\documentstyle[aps,preprint]{revtex}
\begin{document}
\draft
\title{Causal Structure and Degenerate Phase Boundaries}
\author{Yongge Ma}
\address{Department of Physics, Beijing Normal University, Beijing 100875, China.}
\author{Canbin Liang}
\address{Center of Theoretical Physics, CCAST (World Laboratory), Beijing 100080,\\
China,\\
and Department of Physics, Beijing Normal University, Beijing 100875, China.%
\thanks{%
Mailing address}}
\maketitle

\begin{abstract}
Time-like and null hypersurfaces in the degenerate space-times in Ashtekar
theory are defined in the light of the degenerate causal structure proposed
by Matschull. Using the new definition of null hypersurfaces, the conjecture
that the ``phase boundary'' separating the degenerate space-time region from
the nondegenerate one in Ashtekar's gravity is always null is proved under
certain circumstances.

\end{abstract}

\newpage 

\section{Introduction}

Ashtekar's formalism of general relativity[1] has led to a considerable
progress in loop quantum gravity[2]. A special feature of this framework is
that degenerate triads, and hence degenerate metrics, are admitted, and the
degenerate metrics play an important role in the quantum description of
gravity[3,4]. The significance of understanding degenerate metrics was
emphasized in Refs.5 and 6. Various kinds of degenerate solutions to
classical Ashtekar's equations have been studied[5-11], and the local causal
structure of degenerate Ashtekar theory has also been established[12]. Using
a ``covariant approach'', Bengtsson and Jacobson[6] investigated the
structure of the ``phase boundaries'' between degenerate and nondegenerate
space-time regions, and conjectured that the phase boundaries should always
be null provided that the metric is a ``regular'' solution to Ashtekar's
equations, that is, solutions in which the canonical variables $(A_i^I,%
\tilde{e}_I^i)$ , the shift vector $N^i$, and the lapse density, $\underline{%
N}$ (weight $-1$), all take finite value which, except for $\underline{N}$,
are allowed to vanish. In a recent paper[13], however, a degenerate phase
boundary is distinguished from its image, and moreover, it is shown that the
definition of the nullness of the image of the phase boundary used in Ref.6
could not be generalized to the phase boundary itself. The main focus of the
present paper, on the other hand, is first to give a reasonable definition
of the nullness of the boundary, and then to prove the conjecture under
certain circumstances.

The suggestion of Ref.6 to create a space-time with a degenerate region by
the covariant approach is as follows. Start off with a nondegenerate metric
which solves Einstein's equations and reparametrize one of the coordinates.
This reparametrization is chosen so that it is not a diffeomorphism at some
particular value of the coordinate. Adopting the new coordinate, the
solution can be smoothly matched to a solution to the Ashtekar equations
with a degenerate metric at the surface where the transformation misbehaves.
To make things clearer we reformulate this procedure as follows. Let $M$ be
a 4-dimensional manifold and $M_1$ a 4-dimensional submanifold with a
3-dimensional boundary $\partial M_1$. Suppose $\hat{M}$ is a 4-dimensional
manifold with a nondegenerate metric $\hat{g}_{\mu \nu }$ which solves the
Einstein's equations, and $\phi $ is a diffeomorphism from $M_1$ to some
open set $\hat{M}_1\subset \hat{M}$. Extend the domain of $\phi $ smoothly
to the whole of $M$ so that $M-M_1$ is mapped onto $\phi [\partial M_1]$,
and the pushforward $\phi _{*}$ restricted to the tangent bundle of $%
\partial $$M_1$ to that of $\phi [\partial M_1]$ is nondegenerate. Then the
pullback $g_{\mu \nu }\equiv \phi ^{*}\hat{g}_{\mu \nu }$ is nondegenerate
on $M_1$ and degenerate on $M-M_1$. One therefore has a space-time $%
(M,g_{\mu \nu })$ with a ``phase boundary'', $\partial M_1$, separating a
nondegenerate region from a degenerate one. It is clear that the
``reparametrization procedure'' mentioned above is a special case of this
treatment.

Inspired by Ref.12, we try to define timelike and null hypersurfaces in a
degenerate space-time in Sec.2. Armed with this new definition for a null
hypersurface, we then give a proof of the conjecture that the phase boundary
is null in Sec.3 under the circumstances where the degenerate space-time
with a phase boundary is obtained by the reparametrization procedure
mentioned above.

\section{ Defining null hypersurfaces in degenerate space-times}

Suppose the boundary $\phi [\partial M_1]$ is given by $f=0$, where $f$ is a
smooth function on $\hat{M}$ with $\nabla _\mu f|_{\phi [\partial M_1]}\neq
0 $ , then $\phi [\partial M_1]$ is defined in Ref.[6] to be null if $g^{\mu
\nu }\nabla _\mu f\nabla _\nu f\rightarrow 0$ as $\phi [\partial M_1]$ ``is
approached from the nondegenerate side''. However, as pointed out in
Ref.[13], this definition is inappropriate to $\partial M_1$ since it
depends upon the choice of the function $f$ on $M$, and concrete examples
show that there exist functions $f$ and $\bar{f}$ with $\lim g^{\mu \nu
}\nabla _\mu f\nabla _\nu f=0$ while $\lim g^{\mu \nu }\nabla _\mu \bar{f}%
\nabla _\nu \bar{f}\neq 0$. This obstacle could be overcome if we use $\sqrt{%
-g}g^{\mu \nu }\nabla _\mu f\nabla _\nu f$ instead of $g^{\mu \nu }\nabla
_\mu f\nabla _\nu f$. In Ashtekar theory there is a well-defined densitized
inverse metric, $\tilde{g}^{\mu \nu }$, with components in any coordinate
system of a $3+1$ decomposition[12]:

\begin{equation}
\tilde{g}^{\mu \nu }=\left( 
\begin{array}{cc}
\tilde{g}^{tt} & \tilde{g}^{ti} \\ 
\tilde{g}^{jt} & \tilde{g}^{ji}
\end{array}
\right) =\left( 
\begin{array}{cc}
-\underline{N}^{-1} & \underline{N}^{-1}N^i \\ 
\underline{N}^{-1}N^j & \ \underline{N}\tilde{h}^{ji}-\underline{N}%
^{-1}N^jN^i
\end{array}
\right) ,
\end{equation}
where $\underline{N}$ and $N^i$ are respectively the lapse density and the
shift vector, and $\tilde{h}^{ji}$ is the densitized inverse 3-metric of
weight $+2$, and in the nondegenerate case one has $\tilde{g}^{\mu \nu }=%
\sqrt{-g}g^{\mu \nu }$. Eq.(1) implies that $\tilde{g}^{\mu \nu }$ remains
finite in the Ashtekar theory of degenerate space-times, and therefore $\lim 
\sqrt{-g}g^{\mu \nu }\nabla _\mu f\nabla _\nu f=\tilde{g}^{\mu \nu }\nabla
_\mu f\nabla _\nu f|_{f=0}.$ Let $f$ and $\bar{f}$ be two distinct functions
on $M$ with $f|_{\partial M_1}=\bar{f}|_{\partial M_1}=0$ and $\nabla _\mu
f|_{f=0}\neq 0$ and $\nabla _\mu \bar{f}|_{\bar{f}=0}\neq 0$, then there
exists a function $\lambda $ on $M$ such that $\nabla _\mu f|_{f=0}=\lambda
\nabla _\mu \bar{f}|_{\bar{f}=0}$, and hence 
\[
\tilde{g}^{\mu \nu }\nabla _\mu f\nabla _\nu f|_{f=0}=\lambda ^2\tilde{g}%
^{\mu \nu }\nabla _\mu \bar{f}\nabla _\nu \bar{f}|_{\bar{f}=0}. 
\]
Hence, $\tilde{g}^{\mu \nu }\nabla _\mu f\nabla _\nu f|_{f=0}=0$ if and only
if $\tilde{g}^{\mu \nu }\nabla _\mu \bar{f}\nabla _\nu \bar{f}|_{\bar{f}%
=0}=0 $. We therefore obtain a self-consistent definition of null
hypersurfaces in a degenerate space-time in Ashtekar's theory as follows:

{\bf Definition }${\bf 1}${\bf :} A hypersurface described by $f=0$ with $%
\nabla _\mu f|_{f=0}\neq 0$ in space-time $(M,g_{\mu \nu })$ is said to be
null if $\tilde{g}^{\mu \nu }\nabla _\mu f\nabla _\nu f|_{f=0}=0$.

In the following we will use the symbol, $\tilde{E}_A^\mu $, to denote the
vierbein of vector densities weighted $+1/2$, i.e., the square roots of $%
\tilde{g}^{\mu \nu }$, namely,

\[
\tilde{g}^{\mu \nu }=\eta ^{AB}\tilde{E}_A^\mu \tilde{E}_B^\nu , 
\]
where $\eta ^{AB}$ is the Minkowski metric to raise (and $\eta _{AB}$ to
lower) the interior indices ``$A$'' and ``$B$''. Note that there is $SO(3,1)$
gauge freedom for $\tilde{E}_A^\mu $, and the components of certain choice
of $\tilde{E}_A^\mu $ in any coordinate system associated with a $3+1$
decomposition are

\begin{equation}
\tilde{E}_A^\mu =\left( 
\begin{array}{cc}
\tilde{E}_0^t & \tilde{E}_0^i \\ 
\tilde{E}_I^t & \tilde{E}_I^i
\end{array}
\right) =\left( 
\begin{array}{cc}
\sqrt{\underline{N}^{-1}} & \ -\sqrt{\underline{N}^{-1}}N^i \\ 
0 & -\sqrt{\underline{N}}\tilde{e}_I^i
\end{array}
\right) ,
\end{equation}
where $\tilde{e}_I^i$ is the densitized triad of weight $+1$ in Ashtekar
theory (with ``$i$'' and ``$I$'' the spatial and interior indices
respectively), and the columns of $(\tilde{E}_A^\mu )$ are labelled by
space-time indices. Given a vierbein, one may consider $\tilde{E}_A^\mu (x):%
{\rm M}^4\rightarrow T_xM$ as a map from the 4-dimensional Minkowski space
into the tangent bundle of the manifold $M$. The ``future'', ${\cal F}(x)$,
of a point, $x\in M$, can therefore be defined[12] as the set of all tangent
vectors at $x$ which are images of some vectors, $\varsigma ^A$, in ${\rm M}%
^4$ satisfying $\varsigma ^A\varsigma _A\leq 0$ and $\varsigma ^0>0$ , i.e.,

\[
{\cal F}(x)\equiv \{v^\mu (x)\in T_xM\ |\ \exists \ \varsigma ^A\in {\rm M}%
^4,\varsigma ^A\varsigma _A\leq 0,\varsigma ^0>0,\ such\ that\ v^\mu
(x)=\varsigma ^A\tilde{E}_A^\mu (x)\}. 
\]
Thus, depending on the rank of the vierbein, the future, ${\cal F}(x)$, is
either a (4-dimensional) hypercone ($rank\ \tilde{E}_A^\mu =4$), a
(3-dimensional) cone ($rank\ \tilde{E}_A^\mu =3$), an angle ($rank\ \tilde{E}%
_A^\mu =2$), or a half-line ($rank\ \tilde{E}_A^\mu =1$) in Ashtekar theory.
This local causal structure can be used to define the timelike and null
hypersurfaces as follows.

{\bf Definition }${\bf 2}${\bf :} A hypersurface $\Sigma $ is said to be
timelike if for any point $x\in \Sigma $ the tangent space, $T_x\Sigma $
(tangent to $\Sigma $), of $x$ contains a nonzero vector, $v^\mu $, which is
the image under the mapping $\tilde{E}_A^\mu $ of a timelike vector, $%
\varsigma ^A$, in the Minkowski space, i.e., $\exists \ v^\mu =\varsigma ^A%
\tilde{E}_A^\mu \in T_x\Sigma $ such that $\eta _{AB}\varsigma ^A\varsigma
^B<0$.

{\bf Definition }${\bf 3}${\bf : }A hypersurface $\Sigma $ is said to be
null if for any point $x\in \Sigma $ the tangent space, $T_x\Sigma ,$ of $x$
contains a nonzero vector that is the image of a null vector in the
Minkowski space, and there exists no timelike vector, $\varsigma ^A$, in the
Minkowski space such that $\varsigma ^A\tilde{E}_A^\mu \in T_x\Sigma $.

Definitions $2$ and $3$ are consistent with the causal structure and can be
re-formulated in terms of ${\cal F}(x)$ as follows: Let $i({\cal F}(x))$ and 
$\partial {\cal F}(x)$ represent respectively the interior and the boundary
of ${\cal F}(x)$, then a hypersurface $\Sigma $ is timelike if and only if $%
T_x\Sigma \cap i({\cal F}(x))\neq \emptyset $, while $\Sigma $ is null if
and only if $T_x\Sigma \cap {\cal F}(x)\neq \emptyset $ and $T_x\Sigma \cap 
{\cal F}(x)\subset \partial {\cal F}(x)$. Note that the definition of a
spacelike hypersurface $\Sigma $ given by Ref.12 is equivalent to requiring $%
T_x\Sigma \cap {\cal F}(x)=\emptyset $. Note also that both Definitions $2$
and $3$ are applicable to the cases where the ranks of $\tilde{E}_A^\mu $
are two, three, and four. In the case where $\tilde{E}_A^\mu $ is of rank
one, the timelike and null hypersurfaces become indistinguishable.

Now it is natural to ask whether Definition $3$ is equivalent to Definition $%
1$. Suppose a hypersurface defined by $f=0$ with $\nabla _\mu f|_{f=0}\neq 0$
is null according to Definition $3$, then any vector field, $v^\mu $,
tangent to the hypersurface satisfies

\begin{equation}
0=v^\mu \nabla _\mu f=\varsigma ^A\tilde{E}_A^\mu \nabla _\mu f=\varsigma
^A\omega _A,
\end{equation}
where $\varsigma ^A$ is any inverse image of $v^\mu $ under $\tilde{E}_A^\mu 
$, and $\omega _A\equiv \tilde{E}_A^\mu \nabla _\mu f$. Since $\varsigma ^A$
can be null but not timelike, it follows from Eq.(3) that $\omega ^B\equiv
\eta ^{BA}\omega _A$ must be a null vector. Consequently on $f=0$ we have 
\[
\tilde{g}^{\mu \nu }\nabla _\mu f\nabla _\nu f=\eta ^{AB}\tilde{E}_A^\mu 
\tilde{E}_B^\nu \nabla _\mu f\nabla _\nu f=\eta ^{AB}\omega _A\omega _B=0, 
\]
i.e., the hypersurface $f=0$ is also null according to Definition $1$.
However, the degeneracy of $\tilde{E}_A^\mu $ implies the possibility of $%
\omega _A\equiv \tilde{E}_A^\mu \nabla _\mu f=0$ , in this case the
hypersurface is null according to Definition $1$ while might well be nonnull
according to Definition $3$.

The above arguments lead to the following equivalent definition of
Definition $3$:

{\bf Definition }${\bf 3}^{\prime }${\bf :} A hypersurface $f=0$ (with $%
\nabla _\mu f|_{f=0}\neq 0$) is null if $\omega _A\equiv \tilde{E}_A^\mu
\nabla _\mu f|_{f=0}$ is a nonzero null covector in the Minkowski space.

Since Definition $3^{\prime }$ is consistent with the local causal structure
and convenient to use, we will use it to judge whether the phase boundary $%
\partial M_1$ is null in the next section.

It should be noted that the choice of the gauge as well as the coordinate
system for $\tilde{E}_A^\mu $ is irrelevant. The interior gauge
transformation preserves the Minkowski metric and hence does no harm to the
previous discussions. Since $\tilde{E}_A^\mu $ are vector densities, for a
vector $\varsigma ^A$ in Minkowski space, the image $v^\mu (x)\equiv \tilde{E%
}_A^\mu \varsigma ^A$ , viewed as a vector at $x\in M$, will change under a
coordinate transformation to $v^{\prime \mu }(x)\equiv \tilde{E}_A^{\prime
\mu }\varsigma ^A$. However, the transformation law for the components of a
vector density guarantees that $v^{\prime \mu }(x)$ and $v^\mu (x)$ have the
same direction, therefore coordinate transformations do no harm to the
previous discussions either.

\section{ Nullness of the degenerate phase boundary $\partial M_1$}

We assume in this section that the degenerate phase boundary is obtained
through the covariant approach mentioned in Sec.1. As shown in Ref.13, the
hypersurface $\phi [\partial M_1]$ in $\hat{M}$ must be null if the pullback
metric $g_{\mu \nu }$ on $M$ is to be a regular solution to Ashtekar's
equations. The argument is as follows in short. In any $3+1$ decomposition
of the space-time $(M,g_{\mu \nu })$ one has $h_{ij}N^j=g_{0i}\ (i=1,2,3)$,
where $h_{ij}$ and $N^j$ are respectively the 3-metric and the shift vector.
Since $h\equiv \det (h_{ij})=0$ in $M-M_1$, the last three columns of $%
(g_{\mu \nu })$ must be linearly dependent to ensure the finiteness of $N^i$%
. Hence there exists a 4-vector $T^\nu =(0,\lambda ^i)$ at each point of $%
M-M_1$ with $\lambda ^i$ a non-vanishing 3-vector such that $g_{\mu \nu
}T^\nu =0$. Furthermore, the lapse scalar $N$ must vanish to keep the lapse
density $\underline{N}$ finite in $M-M_1$, it then follows from $%
g_{00}=-N^2+g_{0i}N^i$ that there exists another 4-vector $S^\nu =(1,-N^i)$
at each point of $M-M_1$ such that $g_{\mu \nu }S^\nu =0$. Therefore $T^\nu $
and $S^\nu $ represent two independent degenerate directions of $g_{\mu \nu
} $, and hence there must be some degenerate vector field, $W^\nu $, that is
tangent to $\partial M_1$. It follows from the nondegeneracy of the
pushforward $\phi _{*}$ (restricted to $\partial M_1$ ) that there is a
vector field, $\phi _{*}W^\nu $, on $\phi [\partial M_1]$ that is the
desired null normal to $\phi [\partial M_1]$. It is also clear from this
argument that the rank of $g_{\mu \nu }$ is two on $\partial M_1$, from
which one can argue that $rank(\tilde{g}^{\mu \nu })=2$, and hence $rank(%
\tilde{E}_A^\mu )$ must be two or three on $\partial M_1$. We could
therefore use Definition $3^{\prime }$ in Sec.2 to judge whether $\partial
M_1$ is null.

Without loss of generality, we choose a ``time orthogonal'' $3+1$
decomposition of the space-time $(\hat{M},\hat{g}_{\mu \nu })$, and the line
element reads

\begin{equation}
d\hat{s}^2=-\hat{N}^2dT^2+\hat{h}_{ij}dX^idX^j.
\end{equation}
Let $U$ be a smooth function on $M$ with $\nabla _\mu U\neq 0$ and $U=0$
represents $\phi [\partial M_1]$, then

\begin{equation}
dU=\hat{\beta}(T,X^j)dT+\hat{\alpha}_i(T,X^j)dX^i,
\end{equation}
where $\hat{\beta}\equiv \partial U/\partial T$ and $\hat{\alpha}_i\equiv
\partial U/\partial X^i,i=1,2,3$. It follows from Eqs.(4), (5) and the
nullness of $\phi [\partial M_1]$, i.e., $\hat{g}^{\mu \nu }\nabla _\mu
U\nabla _\nu U|_{U=0}=0$ that

\begin{equation}
\lbrack \hat{h}^{ij}\hat{\alpha}_i\hat{\alpha}_j-(\hat{\beta}/\hat{N}%
)^2]|_{U=0}=0,
\end{equation}
where $\hat{h}^{ij}$ is the inverse of the 3-metric $\hat{h}_{ij}$, and $%
\hat{N}$ is the lapse scalar. The line element (4) in the domain of $U$ can
be re-expressed as

\begin{equation}
d\hat{s}^2=(\hat{N}/\hat{\beta})^2[-dU^2+2\hat{\alpha}_idX^idU-(\hat{\alpha}%
_idX^i)^2]+\hat{h}_{ij}dX^idX^j.
\end{equation}
The mapping $\phi :M\rightarrow \hat{M}$ induces four functions $\phi
^{*}U,\phi ^{*}X^i(i=1,2,3)$ on $M$ with $\phi ^{*}U|_{M-M_1}=0$. Without
essential loss of generality, let $(u,x^i)$ be a local coordinate system on $%
M$ covering a neighborhood of $\partial M_1$ with $u|_{\partial
M_1}=0,u|_{M_1}>0$ [hereafter $M_1$ (or $M$) is short for ``the interaction
of $M_1$ (or $M$) and the coordinate patch''] and $x^i=\phi ^{*}X^i$, then
one has a function $U(u)$ [short for $(\phi ^{*}U)(u)$ ] with $U^{\prime
}(u)|_{M-M_1}\equiv [dU/du]_{M-M_1}=0$. It then follows from Eq.(7) that the
line element of $g_{\mu \nu }\equiv \phi ^{*}\hat{g}_{\mu \nu }$ in this
coordinate system reads

\begin{equation}
ds^2=(\hat{N}/\hat{\beta})^2[-(U^{\prime })^2du^2+2U^{\prime }\hat{\alpha}%
_idx^idu-(\hat{\alpha}_idx^i)^2]+\hat{h}_{ij}dx^idx^j.
\end{equation}
Let $u=u(t,x^i)$, where $t$ is the time coordinate of certain $3+1$
decomposition of $(M,g_{\mu \nu })$, and

\[
du=\beta (t,x^i)dt+\alpha _i(t,x^i)dx^i, 
\]
where $\beta \equiv \partial u/\partial t$ and $\alpha _i\equiv \partial
u/\partial x^i,i=1,2,3$, then the $3+1$ decomposition of the metric (8) reads

\begin{equation}
ds^2=-(\hat{N}/\hat{\beta})^2[(U^{\prime }\beta )^2dt^2+2U^{\prime }\beta
\gamma _idx^idt+(\gamma _idx^i)^2]+\hat{h}_{ij}dx^idx^j,
\end{equation}
where $\gamma _i\equiv U^{\prime }\alpha _i-\hat{\alpha}_i$. The
determinant, $g$, of the line element (9), the spatial 3-metric, $h_{ij}$,
induced by metric (9), and the determinant, $h$ , of $h_{ij}$ can be
obtained through straightforward calculations as

\begin{equation}
g\equiv \det \left( g_{\mu \nu }\right) =-(U^{\prime }\beta \hat{N}/\hat{%
\beta})^2\hat{h},\quad \hat{h}\equiv \det \left( \hat{h}_{ij}\right) ,
\end{equation}

\begin{equation}
h_{ij}=\hat{h}_{ij}-(\hat{N}/\hat{\beta})^2\gamma _i\gamma _j,
\end{equation}

\begin{equation}
h\equiv \det \left( h_{ij}\right) =\hat{h}[1-(\hat{N}/\hat{\beta})^2\gamma
_i\gamma _j\hat{h}^{ij}].
\end{equation}
Since $g=-N^2h$, where $N$ is the lapse scalar, it follows from Eqs.(10) and
(12) that the Ashtekar's lapse density on $M_1$ is

\begin{equation}
\underline{N}=\frac N{\sqrt{h}}=\frac{\sqrt{-g}}h=\frac{\left| \beta \hat{%
\beta}U^{\prime }\right| }{\hat{N}\sqrt{\hat{h}}[(\hat{\beta}/\hat{N}%
)^2-\gamma _i\gamma _j\hat{h}^{ij}]}.
\end{equation}
Since the shift vector, $N^i$, relates to the metric components of Eq.(9) via

\[
h_{ij}N^j=g_{oi}=-(\hat{N}/\hat{\beta})^2\beta U^{\prime }\gamma _i, 
\]
a straightforward calculation shows that the shift vector on $M_1$ reads

\begin{equation}
N^i=g_{0i}h^{ij}=\frac{U^{\prime }\beta \gamma _j\hat{h}^{ij}}{\hat{h}%
^{lm}\gamma _l\gamma _m-(\hat{\beta}/\hat{N})^2}.
\end{equation}
Using Eq.(6), it is not difficult to show that Eqs.(13) and (14) imply

\begin{equation}
\underline{N}|_{\partial M_1}=\lim_{u\rightarrow 0^{+}}\frac{\left| \beta 
\hat{\beta}U^{\prime }\right| }{\hat{N}\sqrt{\hat{h}}[(\hat{\beta}/\hat{N}%
)^2-\gamma _i\gamma _j\hat{h}^{ij}]}=\frac{\left| \beta \hat{\beta}\right| }{%
\hat{N}\sqrt{\hat{h}}\left| 2\hat{\alpha}_i\alpha _j\hat{h}^{ij}+B\right| },
\end{equation}
and

\begin{equation}
N^i|_{\partial M_1}=\lim_{u\rightarrow 0^{+}}\frac{U^{\prime }\beta \gamma _j%
\hat{h}^{ij}}{\hat{h}^{lm}\gamma _l\gamma _m-(\hat{\beta}/\hat{N})^2}=\frac{%
\beta \hat{\alpha}_j\hat{h}^{ij}}{2\hat{h}^{lm}\hat{\alpha}_l\alpha _m+B},
\end{equation}
where

\begin{equation}
B\equiv \lim_{u\rightarrow 0^{+}}\frac 1{U^{\prime }}[(\hat{\beta}/\hat{N}%
)^2-\hat{h}^{ij}\hat{\alpha}_i\hat{\alpha}_j].
\end{equation}
If we choose the vierbein, $\tilde{E}_A^\mu $, as Eq.(2), then, according to
Definition $3^{\prime }$, the key quantity for judging whether the
hypersurface $u=const.$ is null is

\begin{equation}
u_A\equiv \tilde{E}_A^\mu (du)_\mu =\left( 
\begin{array}{c}
\sqrt{\underline{N}^{-1}}(\beta -\alpha _iN^i) \\ 
-\sqrt{\underline{N}}\alpha _i\tilde{e}_I^i
\end{array}
\right) \equiv \left( 
\begin{array}{c}
u_0 \\ 
u_I
\end{array}
\right) ,\quad I=1,2,3,
\end{equation}
where

\[
(du)_\mu =\left( 
\begin{array}{c}
\beta \\ 
\alpha _i
\end{array}
\right) ,\quad i=1,2,3. 
\]
It follows from Eqs.(15), (16), and (18) that

\begin{equation}
(u_0)^2|_{\partial M_1}=\left| \frac \beta {\hat{\beta}}\right| \frac{\hat{N}%
\sqrt{\hat{h}}(\hat{\alpha}_i\alpha _j\hat{h}^{ij}+B)^2}{\left| 2\hat{h}^{lm}%
\hat{\alpha}_l\alpha _m+B\right| },
\end{equation}
and

\begin{equation}
(u_1)^2+(u_2)^2+(u_3)^2=\underline{N}\alpha _i\alpha _j\tilde{e}_I^i\tilde{e}%
^{Ii}=\underline{N}\alpha _i\alpha _j\tilde{h}^{ij}.
\end{equation}
Through a non-trivial calculation, which will be given in the Appendix, we
get

\begin{equation}
\alpha _i\alpha _j\tilde{h}^{ij}|_{\partial M_1}=\hat{h}(\hat{N}/\hat{\beta}%
)^2(\hat{\alpha}_i\alpha _j\hat{h}^{ij})^2.
\end{equation}
Hence Eq.(20) evaluated on $\partial M_1$ gives

\begin{equation}
\lbrack (u_1)^2+(u_2)^2+(u_3)^2]|_{\partial M_1}=\left| \frac \beta {\hat{%
\beta}}\right| \frac{\hat{N}\sqrt{\hat{h}}(\hat{\alpha}_i\alpha _j\hat{h}%
^{ij})^2}{\left| 2\hat{h}^{lm}\hat{\alpha}_l\alpha _m+B\right| },
\end{equation}
which equals Eq.(19) if and only if $B=0$. To show that this condition is
indeed satisfied we first obtain from Eq.(17) the following expression of $B$%
:

\begin{equation}
B=\lim_{u\rightarrow 0^{+}}[\frac{\partial (\hat{\beta}^2/\hat{N}^2-\hat{h}%
^{ij}\hat{\alpha}_i\hat{\alpha}_j)}{\partial U}\frac{U^{\prime }}{U^{\prime
\prime }}],
\end{equation}
where $U^{\prime \prime }\equiv dU^{\prime }/du$. If

\begin{equation}
b\equiv \lim_{u\rightarrow 0^{+}}\frac{U^{\prime }}{U^{\prime \prime }}\neq
0,
\end{equation}
there exists a smooth function $L(u)$ on $M$ such that $U^{\prime \prime
}/U^{\prime }=L(u)$ on $M_1$, and $\lim_{u\rightarrow 0^{+}}(U^{\prime
\prime }/U^{\prime })=L(0)=1/b$. Therefore one has

\begin{equation}
U^{\prime }=C\exp [\int_0^uL(\tau )d\tau ]\quad on\ M_1
\end{equation}
where $C=const.$ . Since $\lim_{u\rightarrow 0^{+}}U^{\prime }=0$, Eq.(25)
implies that $C=0$ , and hence $U^{\prime }|_{M_1}=0$, contradicting the
above mentioned requirement of the mapping $\phi $. Therefore the assumption
(24) is false, and we have

\[
\lim_{u\rightarrow 0^{+}}\frac{U^{\prime }}{U^{\prime \prime }}=0, 
\]
and hence it follows from Eq.(23) that $B=0$ since $\partial (\hat{\beta}^2/%
\hat{N}^2-\hat{h}^{ij}\hat{\alpha}_i\hat{\alpha}_j)/\partial U$ is regular
on $M$. Consequently the vector $u^A|_{\partial M_1}$ is null in the
Minkowski space. Furthermore, since $\underline{N}$ should be finite for a
regular solution to Ashtekar's equations, Eq.(15) implies that Eq.(22) does
not vanish, ensuring that $u^A|_{\partial M_1}\neq 0$. We therefore conclude
that the phase boundary, $\partial M_1$, represented by $u=0$, is null.

It is also worth noting that, in contrast with what the authors of Ref.6
argued, (This is originally, in Ref.6, referred to the image rather than the
phase boundary itself.) the Ashtekar's evolution equations are not at all a
necessary condition for the degenerate phase boundary $\partial M_1$ to be
null.

\acknowledgments

The authors would like to thank Prof. Zhiquan Kuang for his enlightening
suggestions and valuable comments. One of the authors, Y. Ma, also would
like to thank Prof. Ted Jacobson for helpfull discussions. This work has
been supported by the National Science Foundation of China.

\appendix 

\section{Proof for Eq.(21)}

Denoting by $A^{ij}$ and $\hat{A}^{ij}$ the complementary minors of $h_{ij}$
and $\hat{h}_{ij}$ respectively, it follows from Eq.(11) that 
\begin{eqnarray}
\alpha _i\alpha _j\tilde{h}^{ij} &=&(-1)^{i+j}\alpha _i\alpha _jA^{ij} 
\nonumber \\
&=&(\alpha _1)^2A^{11}+(\alpha _2)^2A^{22}+(\alpha _3)^2A^{33}+2(-\alpha
_1\alpha _2A^{12}+\alpha _1\alpha _3A^{13}-\alpha _2\alpha _3A^{23}) 
\nonumber \\
&=&(\alpha _1)^2\{\hat{A}^{11}-(\hat{N}/\hat{\beta})^2[\hat{h}_{22}(\hat{%
\alpha}_3)^2+\hat{h}_{33}(\hat{\alpha}_2)^2-2\hat{h}_{23}\hat{\alpha}_2\hat{%
\alpha}_3]\}  \nonumber \\
&&+(\alpha _2)^2\{\hat{A}^{22}-(\hat{N}/\hat{\beta})^2[\hat{h}_{11}(\hat{%
\alpha}_3)^2+\hat{h}_{33}(\hat{\alpha}_1)^2-2\hat{h}_{13}\hat{\alpha}_1\hat{%
\alpha}_3]\}  \nonumber \\
&&+(\alpha _3)^2\{\hat{A}^{33}-(\hat{N}/\hat{\beta})^2[\hat{h}_{22}(\hat{%
\alpha}_1)^2+\hat{h}_{11}(\hat{\alpha}_2)^2-2\hat{h}_{21}\hat{\alpha}_2\hat{%
\alpha}_1]\}  \nonumber \\
&&+2\alpha _1\alpha _2\{-\hat{A}^{12}+(\hat{N}/\hat{\beta})^2[\hat{h}_{21}(%
\hat{\alpha}_3)^2+\hat{h}_{33}\hat{\alpha}_1\hat{\alpha}_2-\hat{h}_{13}\hat{%
\alpha}_2\hat{\alpha}_3-\hat{h}_{23}\hat{\alpha}_1\hat{\alpha}_3]\} 
\nonumber \\
&&+2\alpha _3\alpha _2\{-\hat{A}^{32}+(\hat{N}/\hat{\beta})^2[\hat{h}_{23}(%
\hat{\alpha}_1)^2+\hat{h}_{11}\hat{\alpha}_3\hat{\alpha}_2-\hat{h}_{13}\hat{%
\alpha}_2\hat{\alpha}_1-\hat{h}_{21}\hat{\alpha}_1\hat{\alpha}_3]\} 
\nonumber \\
&&+2\alpha _1\alpha _3\{\hat{A}^{13}-(\hat{N}/\hat{\beta})^2[\hat{h}_{21}%
\hat{\alpha}_2\hat{\alpha}_3+\hat{h}_{32}\hat{\alpha}_1\hat{\alpha}_2-\hat{h}%
_{13}(\hat{\alpha}_2)^2-\hat{h}_{22}\hat{\alpha}_1\hat{\alpha}_3]\} 
\nonumber \\
&=&(-1)^{i+j}\alpha _i\alpha _j\hat{A}^{ij}-(\hat{N}/\hat{\beta})^2[\hat{h}%
_{11}(\alpha _2\hat{\alpha}_3-\alpha _3\hat{\alpha}_2)^2+\hat{h}_{22}(\alpha
_3\hat{\alpha}_1-\alpha _1\hat{\alpha}_3)^2  \nonumber \\
&&+\hat{h}_{33}(\alpha _2\hat{\alpha}_1-\alpha _1\hat{\alpha}_2)^2+2\hat{h}%
_{21}(\alpha _3\hat{\alpha}_2-\alpha _2\hat{\alpha}_3)(\alpha _1\hat{\alpha}%
_3-\alpha _3\hat{\alpha}_1)  \nonumber \\
&&+2\hat{h}_{32}(\alpha _1\hat{\alpha}_2-\alpha _2\hat{\alpha}_1)(\alpha _3%
\hat{\alpha}_1-\alpha _1\hat{\alpha}_{3\frac {}{}})+2\hat{h}_{31}(\alpha _2%
\hat{\alpha}_1-\alpha _1\hat{\alpha}_2)(\alpha _3\hat{\alpha}_2-\alpha _2%
\hat{\alpha}_3)]  \nonumber \\
&=&\hat{h}\hat{h}^{ij}\alpha _i\alpha _j-(\hat{N}/\hat{\beta})^2\hat{h}\{[%
\hat{h}^{22}\hat{h}^{33}-(\hat{h}^{23})^2](\alpha _2\hat{\alpha}_3-\alpha _3%
\hat{\alpha}_2)^2  \nonumber \\
&&+[\hat{h}^{11}\hat{h}^{33}-(\hat{h}^{13})^2](\alpha _3\hat{\alpha}%
_1-\alpha _1\hat{\alpha}_3)^2+[\hat{h}^{22}\hat{h}^{11}-(\hat{h}%
^{21})^2](\alpha _2\hat{\alpha}_1-\alpha _1\hat{\alpha}_2)^2  \nonumber \\
&&+2(\hat{h}^{31}\hat{h}^{23}-\hat{h}^{21}\hat{h}^{33})(\alpha _3\hat{\alpha}%
_2-\alpha _2\hat{\alpha}_3)(\alpha _1\hat{\alpha}_3-\alpha _3\hat{\alpha}_1)
\nonumber \\
&&+2(\hat{h}^{31}\hat{h}^{21}-\hat{h}^{11}\hat{h}^{32})(\alpha _1\hat{\alpha}%
_2-\alpha _2\hat{\alpha}_1)(\alpha _3\hat{\alpha}_1-\alpha _1\hat{\alpha}_3)
\nonumber \\
&&+2(\hat{h}^{21}\hat{h}^{23}-\hat{h}^{31}\hat{h}^{22})(\alpha _3\hat{\alpha}%
_2-\alpha _2\hat{\alpha}_3)(\alpha _2\hat{\alpha}_1-\alpha _1\hat{\alpha}%
_2)\}
\end{eqnarray}
Let $H\equiv \hat{h}^{ij}\alpha _i\alpha _j$, then, from Eq.(6) one gets 
\begin{equation}
(\hat{\beta}/\hat{N})^2H|_{\partial M_1}=\hat{h}^{lm}\hat{\alpha}_l\hat{%
\alpha}_m\hat{h}^{ij}\alpha _i\alpha _j.
\end{equation}
Denoting by $L$ the last brace of Eq.(A1), it follows from Eqs.(A1) and (A2)
that 
\begin{eqnarray*}
&& \\
(\frac{\hat{\beta}^2}{\hat{h}\hat{N}^2})\alpha _i\alpha _j\tilde{h}%
^{ij}|_{\partial M_1} &=&\hat{h}^{lm}\hat{\alpha}_l\hat{\alpha}_m\hat{h}%
^{ij}\alpha _i\alpha _j-L \\
&=&(\hat{h}^{11}\alpha _1\hat{\alpha}_1)^2+(\hat{h}^{22}\alpha _2\hat{\alpha}%
_2)^2+(\hat{h}^{33}\alpha _3\hat{\alpha}_3)^2+[\hat{h}^{12}(\alpha _1\hat{%
\alpha}_2+\alpha _2\hat{\alpha}_1)]^2 \\
&&+[\hat{h}^{23}(\alpha _3\hat{\alpha}_2+\alpha _2\hat{\alpha}_3)]^2+[\hat{h}%
^{13}(\alpha _1\hat{\alpha}_3+\alpha _3\hat{\alpha}_1)]^2+2[\hat{h}^{11}\hat{%
h}^{33}\alpha _1\alpha _3\hat{\alpha}_1\hat{\alpha}_3 \\
&&+\hat{h}^{11}\hat{h}^{22}\alpha _1\alpha _2\hat{\alpha}_1\hat{\alpha}_2+%
\hat{h}^{22}\hat{h}^{33}\alpha _2\alpha _3\hat{\alpha}_2\hat{\alpha}_3+\hat{h%
}^{12}\hat{h}^{23}(\alpha _1\hat{\alpha}_2+\alpha _2\hat{\alpha}_1)(\alpha _3%
\hat{\alpha}_2+\alpha _2\hat{\alpha}_3) \\
&&+\hat{h}^{12}\hat{h}^{13}(\alpha _1\hat{\alpha}_2+\alpha _2\hat{\alpha}%
_1)(\alpha _3\hat{\alpha}_1+\alpha _1\hat{\alpha}_3)+\hat{h}^{31}\hat{h}%
^{23}(\alpha _1\hat{\alpha}_3+\alpha _3\hat{\alpha}_1)(\alpha _3\hat{\alpha}%
_2+\alpha _2\hat{\alpha}_3)] \\
&&+2(\hat{h}^{11}\alpha _1\hat{\alpha}_1+\hat{h}^{22}\alpha _2\hat{\alpha}_2+%
\hat{h}^{33}\alpha _3\hat{\alpha}_3)[\hat{h}^{12}(\alpha _1\hat{\alpha}%
_2+\alpha _2\hat{\alpha}_1)+\hat{h}^{23}(\alpha _3\hat{\alpha}_2+\alpha _2%
\hat{\alpha}_3) \\
&&+\hat{h}^{13}(\alpha _1\hat{\alpha}_3+\alpha _3\hat{\alpha}_1)] \\
&=&(\hat{h}^{ij}\alpha _i\hat{\alpha}_j)^2,
\end{eqnarray*}
and hence 
\[
\alpha _i\alpha _j\tilde{h}^{ij}|_{\partial M_1}=\hat{h}(\hat{N}/\hat{\beta}%
)^2(\hat{\alpha}_i\alpha _j\hat{h}^{ij})^2\text{.} 
\]

\end{document}